\documentclass[preprint]{revtex4} 

\usepackage{amssymb, amsmath}
\usepackage[dvips,final]{graphicx}
\usepackage[english]{babel}
\usepackage{bm}
\usepackage{color}

\newcommand{\ndg}{{\phantom{\dagger}}}
\newcommand{\dg}{\dagger}

\newcommand{\ew}[1]{{\langle} #1 {\rangle}}

\newcommand{\figuresize}{8cm}
\newcommand{\dt}{\text{d}t\ }
\newcommand{\dk}{\text{d}k\ }

\newcommand{\ket}[1]{\left| #1 \right\rangle}

\newcommand{\ketbra}[2]{\left| #1 \right\rangle \left\langle #2 \right| }

\begin{document}
\author{Julia Kabuss, Florian Katsch, Andreas Knorr, and Alexander Carmele}
\affiliation{Nichtlineare Optik und Quantenelektronik, Institut f\"ur 
Theoretische Physik, Technische Universit\"at Berlin, Germany}
\title{Unraveling coherent quantum feedback for Pyragas control}
\begin{abstract}
We present a Heisenberg operator based formulation of coherent quantum 
feedback and Pyragas control.
This model is easy to implement and allows for an efficient 
and fast calculation of the dynamics of feedback-driven observables
as the number of contributing correlations grows in systems with 
a fixed number of excitations only linearly in time.
Furthermore, our model unravels the quantum kinetics of entanglement
growth in the system by explicitly calculating non-Markovian multi-time
correlations, e.g., how the emission of a photon is 
correlated with an absorption process in the past.
Therefore, the time-delayed differential equations are expressed 
in terms of insightful physical quantities.
Another considerate advantage of this method is its 
compatibility to typical approximation schemes, such as factorization 
techniques and the semi-classical treatment of coherent fields. 
This allows the application on a variety of setups, ranging from
closed quantum systems in the few excitation regimes to open systems
and Pyragas control in general.
\end{abstract}
\maketitle
\date{\today}
\selectlanguage{english}

\begin{figure}[b!]
\centering
\includegraphics[width=\figuresize]{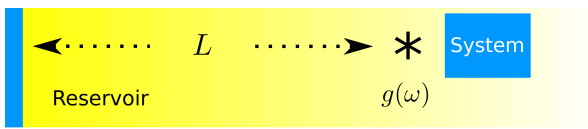}
\caption{We investigate an intrinsic and coherent feedback mechanism, where a 
system is driven by its own past via a continuum of modes. The roundtrip time 
$\tau=2L/c$ scales with the length between the system and the mirror. The 
feedback strength is determined by the coupling constant $g$.}
\label{fig:system}
\end{figure}

\section{Introduction}\label{sec:introduction}
Controlling quantum systems with respect to their coherence properties,
quantum states and dynamics is a basic requirement for quantum information 
science applications \cite{zoller-roadmap,Wiseman::09,brandes2015feedback}. 
The possibilities range from coherent control of quantum systems 
involving higher order coherence processes 
\cite{kiraz2004quantum,wolters2014deterministic}
to open loop feedback control schemes \cite{Wiseman::09,sayrin2011real}. 
Alternatively, structured continua allow state preserving 
measurement-free coherent control schemes for time-delayed 
self-feedback, i.e. the subsequent interaction of the system 
with former states of itself 
\cite{dorner-half-cavities, 
carmele2013single,grimsmo2015time,pichler2015photonic}. 
Such a feedback mechanism combines the 
advantage of time delayed feedback control with coherent 
quantum control in order to manipulate
distinctive system degrees of freedom. 
Quantum self-feedback mechanisms are often based on a structured
continuum with multi-mode environmental degrees of 
freedom 
\cite{cook1987quantum,alber1992photon,tufarelli2014non,hein2014optical,
hughes2007coupled}. 
In this paper, we develop a Heisenberg-operator technique 
for a convenient implementation of coherent feedback via the 
interaction of a quantum system with
a quasi-continuous bosonic reservoir 
\cite{hetet2011single,albert2011observing,hughes2007coupled}, related to the 
Langevin approach \cite{gardiner-book}.
In the following, the reservoir degrees of freedom are
eliminated in favor of system operators inheriting 
the feedback delay time.
The complexity of the multi-mode reservoir is transfered to 
the handling of multi-time-correlations. 
This approach allows a straight forward treatment
of coherent time delayed quantum feedback, with a drastic 
reduction of numerical effort.
Furthermore, it gives access to the feedback mechanism
as the entanglement growth is expressed in physical meaningful
quantities, e.g. the creation and annihilation of a photon at
two different times.
These time-correlated quantities provide insight and allow
the application of factorization techniques, as the degree of
entanglement is directly accessible.
Factorization such as cluster expansion \cite{kira2011semiconductor} 
and Born approximation are unavoidable if the transient feeback 
regime \cite{albert2011observing,schulze2014feedback} between classical 
\cite{lang1980external,kantner2015delay,wegert2014integrated,
flunkert2013dynamics}
and quantum feedback is under investigation
\cite{hetet2011single,glaetzle2010single}.
The paper is organized as following. 
After this introduction, Sec.~\ref{sec:introduction}, and before 
introducing feedback Heisenberg operators, we provide a short review
of existing exact models, which form the backbone of future developments
and also include the basic ingredients of coherent feedback control, 
Sec.~\ref{sec:schroedinger}.
In Sec.~\ref{sec:heisenberg}, we derive the basic equation of motion
for the Heisenberg operators and provide an analytical solution in 
case of an empty cavity coupling to a structured continuum 
\cite{lei2012quantum}. 
Given the operator dynamics, we derive the model to stabilize Rabi
oscillations inside a cavity and unravel the otherwise hidden
feedback mechanism, before we conclude in Sec.~\ref{sec:conclusion}.

\section{Analytical solutions in the Schr\"odinger 
Picture}\label{sec:schroedinger}
In this section, we give three examples for analytically solvable 
coherent quantum self-feedback models.
These models are an important benchmark for numerical implementations 
and contain the basic features of quantum feedback.
All three examples restrict the system dynamics to the single-excitation
regime and thus describe a linear quantum feedback mechanism.
\newline \ 
Following the experimental realization of a decaying single-atom in front
of a mirror \cite{hetet2011single}, 
an analytical model for this scenario has been provided 
\cite{dorner-half-cavities,cook1987quantum,glaetzle2010single,
kabuss2015analytical}.
The Hamiltonian includes the radiative coupling of the atomic 
two-level system to the photon continuum with a boundary condition,
imposing the feedback mechanism $ (\hbar=1) $:
\begin{align}
H &= \omega_e P^\dg P + \int \dk\left( \omega_k \ d^\dg_k d^\ndg_k + g_k 
P^\dg d^\ndg_k + g^*_k d^\dg_k P \right),
\end{align}
where $ P=\ketbra{g}{e}$ denotes the atomic operators for the 
excited- 
$ \ket{e} $ and ground-state $ \ket{g}$.
The radiative continuum is included via the photonic creation and 
annihilation operators $ d^{(\dg)}_k $ for a photon in the mode 
$ k = \omega/c$ ($c$: the vacuum speed of light).
The coupling between the atom and the radiative continuum is denoted by
$ g(k)=g_0 \sin(kL) $ and includes the mirror imposed boundary condition at
a distance between mirror and atom of $L$ with a strength of $g_0$.
The length defines the feedback roundtrip time with $ \tau=2L/c $.

Assuming for the radiative continuum the vacuum state $ b^\ndg_k 
\ket{\Psi(0)}=0$, the Hilbert space is restricted to a single excitation
either in the atomic or photonic degrees of freedom.
The wave vector of the system reads:
\begin{align}
\ket{\Psi(t)} =& 
c_e(t) \ket{e,\lbrace 0 \rbrace_k}
+
\int \dk 
c^k_g(t) \ket{g,\lbrace 1 \rbrace_k},
\end{align}
with the excitation in the atom or in the photonic continuum, respectively.
After applying the Schr\"odinger equation and formally integrating the 
equation for $ \dot{c}^k_g$, the equation for the coefficient 
for the excited state $ c_e $ reads:
\begin{align}
\dot{c_e}(t) =& -\Gamma c_e(t) + \Gamma_\tau c_e(t-\tau) \Theta(t-\tau), 
\end{align}
with $ \Gamma=\pi g_0^2/c $, 
$ \Gamma_\tau = \Gamma \exp[i\omega_e\tau]$, and $ \Theta(x) $ the 
Heaviside function: $ \Theta(x) =0 $ for $ x\le0 $ and $ \Theta(x) =1 $
for $ x>0 $.
Here, the basic ingredient of Pyragas control 
$K\left[f(t)-f(t-\tau)\right]$ are given as $K$ allows the 
control of periodic orbits such as Rabi oscillations or 
relaxation ocillations \cite{flunkert2011time}.
This differential equation of motion can be solved in the Laplace domain,
yielding the following dynamics:
\begin{align}
c_e(t) =& \sum_{n=0}^\infty \frac{e^{-\Gamma t}}{n!} 
\left( 
\Gamma_\tau e^{\Gamma_\tau \tau} (t-n \tau)
\right)^n \Theta(t-n \tau) .
\end{align}
The same solution is obtained, if instead of an atom a bosonic mode decays such 
as a cavity photon with the following Hamiltonian:
\begin{align}
H &= \omega_0 c^\dg c 
+ \int \dk\left( \omega_k \ d^\dg_k d^\ndg_k + g_k 
c^\dg d^\ndg_k + g^*_k d^\dg_k c \right),
\end{align}
with $c^{(\dg)} $ creating (annhiliating) a cavity photon with frequency
$ \omega_0 $ and $ [c,c^\dg]=1$. 
Replacing in the derivation above $ e \rightarrow 1 $ and 
$ g \rightarrow 0 $, the same solution applies to this situation as well.
This is a consequence of the single-excitation limit, where the qubit
and the bosonic mode cannot be distinguish from each other and quantum 
non-linearities are yet not included.

It is possible to derive an analytical solution even for a combined
model, where an emitter is coupled to a cavity mode and the cavity mode
decays into the radiative continuum.
The Hamiltonian reads:
\begin{align} \notag
H &= \omega_e P^\dg P + \omega_0 c^\dg c + \int \dk \omega_k \ d^\dg_k 
d^\ndg_k \\
& + M \left(P^\dg c + c^\dg P\right) 
+ \int \dk \left( g_k 
c^\dg d^\ndg_k + g^*_k d^\dg_k c \right), \label{eq:Hamilton_JCM_FB}
\end{align}
where the cavity-emitter coupling strength is denoted by $ M $ and the 
full wave vector includes now three states:
The wave vector of the system reads:
\begin{align}
\ket{\Psi(t)} =& 
c_e(t) \ket{e,0,\lbrace 0 \rbrace_k}
+
c_g(t) \ket{g,1,\lbrace 0 \rbrace_k}
+
\int \dk 
c^k_g(t) \ket{g,0,\lbrace 1 \rbrace_k},
\end{align}
with the excitation in the emitter, cavity or in the photonic continuum, 
respectively.
The differential equation for the ground state with one photon 
in the cavity reads:
\begin{align}
\dot{c_g} =& -\Gamma c_g -i M c_e + \Gamma_\tau c_g(t-\tau) \Theta(t-\tau).
\end{align}
Applying the binomial series and the Laplace transformation:  $n!/(s-a)^{n+1}
\rightarrow t^n\exp[at] $, we yield an expression in the time domain, after 
choosing $ M=\Gamma/2 $ to simplify the expression:
\begin{align} \notag
c_g(t)
&=
\frac{i}{2}
\sum_{n=0}^{\infty} n! 2^{n+1} e^{-\Gamma/2(t-n\tau)+i\omega_0n\tau} 
\Theta(t-n\tau) \\
& \sum_{k=0}^{n} \frac{(-1)^{k}}{k!(n-k)!}
\frac{\left[(t-n\tau)\Gamma/2\right]^{n+1+k}}{(n+1+k)!}
\label{eq:jcm_feedback_analytical}.
\end{align} 
These analytical solutions are given to benchmark the numerical implementations.
They contain already interesting features such as modified decaying rates 
\cite{dorner-half-cavities,eschner2001light}, Rabi oscillations stabilization 
\cite{carmele2013single,grimsmo2015time}, 
entangling cavities within a quantum eraser set-up 
\cite{hein2015entanglement} or enhanced photon polarization entanglement 
stemming from a controlled biexciton cascade \cite{hein2014optical}. 
The delay phase and the delay strength provide an interesting new degree
of freedom to manipulate quantum systems in a self-sustained, closed-loop
and non-invasive approach.
To unravel coherent quantum feedback and link the feedback mechanism directly
to observable quantities, we switch now to the Heisenberg picture and express
the dynamics in terms of time-dependent operators instead of wave vector
coefficients.

\section{Quantum Feeback in the Heisenberg picture} 
\label{sec:heisenberg}
We investigate the case of boson-boson coupling to include the
feedback mechanism.
We present a way to include coherent quantum self-feedback consistently
at a operator level, i.e. with the Langevin operator technique 
\cite{gardiner-book,carmele2014opto}.
We derive the necessary equations of motion and benchmark our
method with the analytical solutions from the previous section. 
Hereby, we gain insight into the mechanism that leads e.g. to the
stabilization of the Rabi oscillations.
\subsection{Equation of motion - approach}
We start with the Hamiltonian \eqref{eq:Hamilton_JCM_FB}, where 
the emitter is coupled to a cavity mode and the cavity
mode couples to the radiative continuum, leading to a decay of the
cavity mode as well as to feedback.
First, we solve the bilinear Hamiltonian in the Heisenberg equation of 
motion approach $ -i\dot{A}=[H,A]$ with $ A $ being an
arbitrary, non-explicitly time dependent operator $(\hbar=1)$.
The equation of motion (EOM) for the reservoir operator $d^\ndg_k$ is 
derived via 
the 
Heisenberg equation of motion:
\begin{align}
\dot{d^\ndg_k} = -i\omega_k d^\ndg_k -i g^*_k \ c .
\end{align}
Formally integrating yields:
\begin{align}
d^\ndg_k(t) = 
d_k(0)\ e^{-i\omega_k t}
-i\int_0^t \dt^\prime g^*_k e^{-i\omega_k(t-t^\prime)} \ c(t^\prime) .
\end{align}
So, we can include the reservoir interaction by plugging this solution
into the equation of motion of the system boson operator:
\begin{align}
\dot{c} &= -i\omega_c \ c - iM\ P   -i \int \ \dk \ g_k \ d_k(t^\prime) \\
&= -i\omega_c\ c - iM\ P
-i \int \dk \ d_k(0) e^{-i\omega_kt}g_k 
- \int_0^t \dt^\prime \ c(t^\prime) f(t,t^\prime) \notag.
\end{align}
The function $ f(t,t^\prime) $ includes the structure of the reservoir
and can be evaluated as the operator is independent of the wave number 
$ k $:
\begin{align}
f(t,t^\prime) := \int \dk |g_k|^2 e^{-i\omega_k(t-t^\prime)}.
\end{align}
If $ g_k \equiv g_0 $, the function yields 
$ f(t,t^\prime)=2\pi\delta(t-t^\prime)g_0^2/c=2\Gamma\delta(t-t^\prime)$ 
with the definition for the loss coefficient.
If $ g_k = g_0 \sin(kL) $, we have a structured continuum with a
feedback mechanism and the function reads with $ \tau=2L/c $: 
\begin{align}
f(t,t^\prime)
= \Gamma
\left(
2\delta(t-t^\prime)
-\delta(t-t^\prime-\tau)
-\delta(t-t^\prime+\tau)
\right).
\end{align}
The equation of motion of the bosonic system operator with feedback reads,
having in mind that $ t^\prime \le t $:
\begin{align} 
\dot{c} &= -\left(i\omega_c + \Gamma \right) \ c(t) -i M\ P 
+\Gamma_\tau c(t-\tau) \Theta(t-\tau) 
-i \Delta B(t) \label{eq:eom_photon_jcm} ,
\end{align}
with the noise operator $ \Delta B(t)= \int \dk \ d_k(0) e^{-i\omega_kt}g_k $.
\subsection{Photon-Photon coupling}
Setting the cavity-emitter coupling to zero ($ M=0 $), the excitation
manifolds of the photon-operator decouple and the problem can be solved
analytically.
The solution of the photon operator in \eqref{eq:eom_photon_jcm}, subjected to 
feedback, but not coupled to an emitter reads:
\begin{align}
c(t) =& 
e^{-\Gamma t} 
\left(
c(0)
+
\Gamma_\tau
\int_0^t 
\dt^\prime 
e^{\Gamma t^\prime}
\Theta(t^\prime - \tau) c(t^\prime -\tau)
\right),
\end{align}
where the noise term $ \Delta B(t)$ is omitted. 
This can be justified by assuming a structured continuum initially in the
vacuum state and by keeping throughout the calculation the normal-ordering.
Given this general calculation, we need the initial state of the 
cavity system, e.g. $\ew{c^\dg(0) c(0)} = N$.
For the first time interval ($t_0\in[0,\tau]$), one yields
\begin{align}
\ew{c^\dg(t_0) c(t_0)} 
=& \ew{e^{-\Gamma t_0}c^\dg(0)e^{-\Gamma t_0}c(0)}
= N e^{-2\Gamma t_0},
\end{align}
The integral part of the solution does not contribute as $ t<\tau$.
The two-time correlation in the first time interval reads: 
$ \ew{c^\dg(t_0)c(0)}= N e^{-\Gamma t_0}$.
For the second time interval ($t_1\in[\tau,2\tau]$) one yields,
\begin{align}
\ew{c^\dg(t_1) c(t_1)}
& =
N
e^{-2\Gamma t_1} \\ \notag
& + 
e^{-2\Gamma t_1} 
\text{Re}
\left[
\Gamma_\tau
\int_\tau^{t_1} \dt^\prime e^{\Gamma t^\prime} 
\ew{c^\dg(t^\prime-\tau) c(0)} 
\right] \\ \notag
& +
|\Gamma_\tau|^2
\int_\tau^{t_1} \dt^\prime e^{\Gamma t^\prime-2\Gamma t_1} 
\int_\tau^{t_1} \dt^{\prime\prime} e^{\Gamma t^{\prime\prime}} 
\ew{c^\dg(t^\prime-\tau) c(t^{\prime\prime}-\tau)}, 
\end{align}
now we can use the solution from the time interval before,
as $ \ew{c^\dg(t^\prime-\tau) c(t^{\prime\prime}-\tau)}= N 
\exp[-\Gamma(t^\prime-\tau+t^{\prime\prime}-\tau)] $.
We derive, after formally integrating: 
\begin{align}
\ew{c^\dg(t_1) c(t_1)}
& =
N |\Gamma_\tau|^2 e^{-2\Gamma(t_1-\tau)} (t_1-\tau)^2 
\\ \notag
& +N 2 \text{Re}
\left[
\Gamma_\tau 
\right]
e^{-\Gamma(2t_1-\tau)}
(t_1-\tau) 
+ N e^{-\Gamma 2t_1}. 
\end{align}
This procedure allows by means of simple integration to calculate
all higher moments of the photon-correlations.
However, due to the linear coupling, the photon statistics is
not changed and feedback does not provide more than a excitation
exchange between the cavity and the continuum.
In the next section, we calculate a more complex problem and demonstrate,
that feedback can lead to a stabilization of Rabi oscillations between
an emitter and the cavity mode.
\subsection{Hierarchy problem and scaling properties}
To illustrate the method, we reproduce within the 
Heisenberg picture the solution given in \eqref{eq:jcm_feedback_analytical}.
For this scenario, initially an emitter decays into the cavity mode and 
this cavity mode is damped and then driven by the self-feedback after a
roundtrip time $ \tau $.
The relevant equations of motion are derived for $n$ feedback intervals
and are discussed with respect to the new occurring quantities. 
It will become clear, that the coupling of a single bosonic mode 
to the feedback reservoir constitutes in the single excitation limit 
a very specific scenario, where most
complications arising from the time-ordering 
procedure can be omitted.
The basic set of equations of motion,
involving the photon coherence $c$ and the electronic polarization $P$, 
that are
used to derive the time-correlated dynamics of the system are given by:
\begin{eqnarray}
\label{eq:basics_c}
\dot{c}_j&=&-\Gamma c_j+\Gamma_\tau c_{j+1}\Theta_{j+1} - i M \ P_j,\\
\label{eq:basics_p}
\dot{P}_j&=& 2i M \ P_j^\dagger P_jc_j - i M \ c_j,
\end{eqnarray}
where $c_j\equiv c(t-j\tau)$ and $t_j\equiv t-j\tau$ and
the noise term is omitted again.
The noise term can be omitted, as long the normal-ordering is conserved
and initially a vacuum state for the reservoir is assumed. 
The single excitation limit is specific as the normal-ordering is 
preserved for all times automatically.
%
\begin{figure}[t!]
\centering
\includegraphics[width=\figuresize]{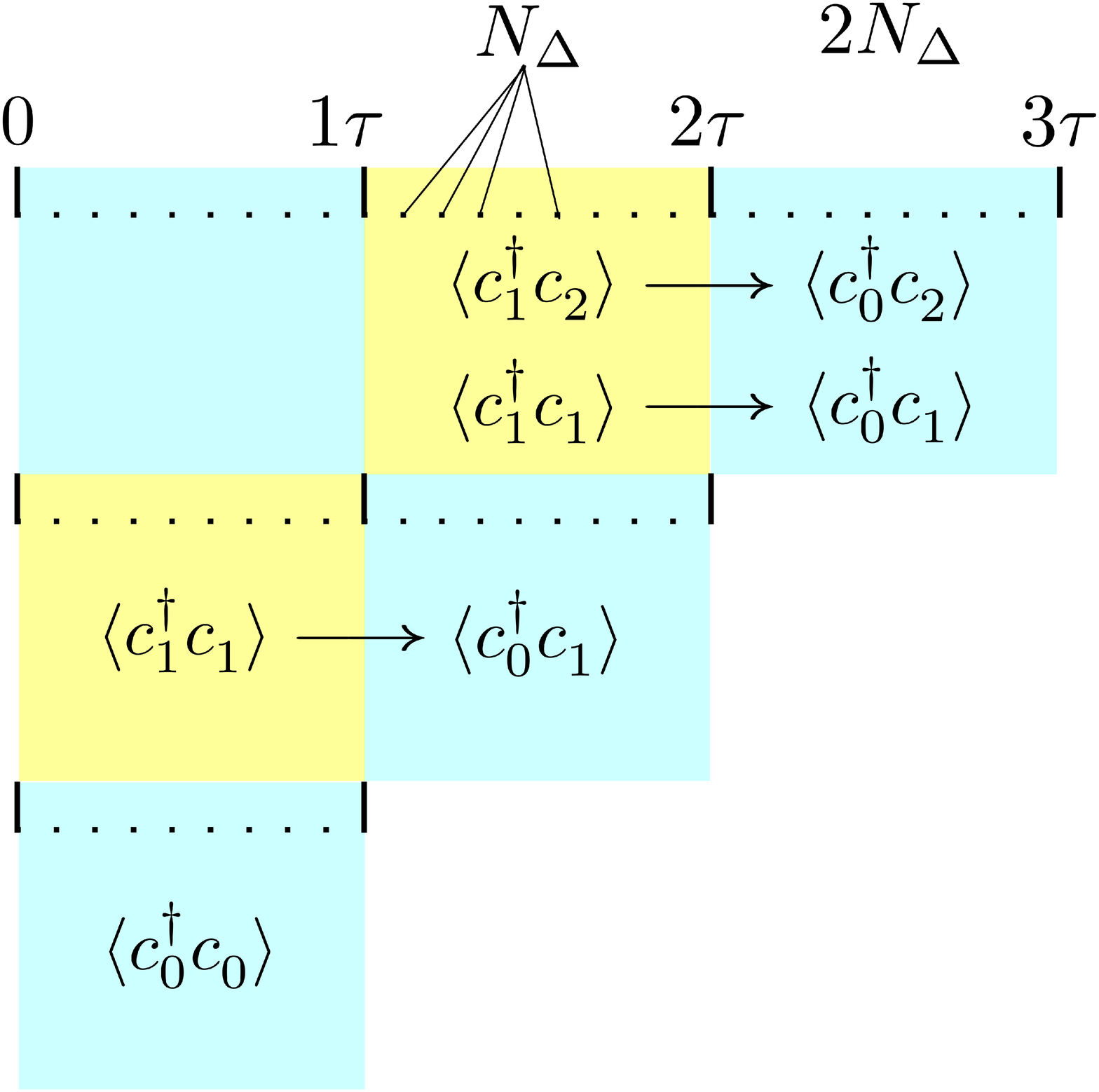}
\caption{Illustration of memory growth: For calculating the dynamics in
$\tau$-interval $i\mathrel{\hat=}[i\tau,(i+1)\tau)$ it is necessary to store
$3iN_\Delta$ two-time correlations. Only information from the previous time
interval $i-1$ is needed. $N_\Delta$ is the number of time discretization
steps.}
\label{fig:linear_growth}
\end{figure}
The electronic degree of freedom $P$ couples only to the cavity mode 
via the electron-photon coupling strength, while the cavity 
coherence $c$ according to \eqref{eq:Hamilton_JCM_FB} is also subject to the 
photon feedback. 
\eqref{eq:basics_c} and \eqref{eq:basics_p} are valid for any time
interval $t_0\in[i\tau,(i+1)\tau)$ with the
advantage, that expectation values of previous times, such as $\langle A_n B_m
\rangle$ have already been calculated on the fly. 
Starting with the $ t=0 $, the equations of motion
correspond to the case of a dissipative Jaynes-Cummings model (JCM):  
$t_0\equiv t\in[0,\tau)$
\begin{eqnarray}
\label{eq:t0_1}
\partial_t\ew{c ^\dagger_0 c^\ndg_0 }
&=&
-2\Gamma\ew{c^\dagger_0 c^\ndg_0 }
+2 \text{Im}\bigl\lbrack M \ \ew{P^\dg_0 c^\ndg_0 }\bigr],\\
\partial_t\ew{P^\dg_0 c^\ndg_0 } 
&=& 
-\Gamma\ew{P^\dg_0 c^\ndg_0 }
+i M \ \ew{c^\dg_0 c^\ndg_0 }
-i M \ \ew{P^\dg_0  P^\ndg_0 },\\
\partial_t\ew{P^\dg_0  P^\ndg_0 }
&=& 
-2\text{Im}\bigl\lbrack  M \  \ew{P^\dg_0 c^\ndg_0 }
\bigr\rbrack. \label{eq:t0_3}
\end{eqnarray}
In the second $\tau$-interval, the EOMs contain additional 
terms in the form of two-time correlated expectation values: 
$t_0\in[\tau,2\tau)$ and $t_1\in[0,\tau)$:
\begin{eqnarray}
\partial_t\ew{c^\dg_0 c^\ndg_0 }
&=&
-2\Gamma\ew{c^\dg_0 c^\ndg_0}
+2\text{Re}\bigl\lbrack  \ew{\Gamma_\tau c^\dg_0 c_1^\ndg}
\bigr\rbrack 
\\ \nonumber 
&+&2\text{Im}\bigl[  M \ \ew{P^\dg_0 c^\ndg_0 }\bigr],\\
\partial_t \ew{P^\dg_0  P^\ndg_0 }
&=& 
-2\text{Im}\bigl\lbrack  M \  \ew{P^\dg_0 c^\ndg_0 } \bigr\rbrack
\end{eqnarray}
\begin{eqnarray}
\partial_t\ew{P^\dg_0 c^\ndg_0 } 
&=& 
-\Gamma\ew{P^\dg_0 c^\ndg_0 }
+\Gamma_\tau\ew{P^\dg_0 c^\ndg_1}\\
\nonumber
&+& i M \ \ew{c^\dg_0 c^\ndg_0 }
-i M \ \ew{P^\dg_0  P^\ndg_0 } ,\\
\partial_t\ew{c^\dg_0 c^\ndg_1}
&=&
-2\Gamma\ew{c^\dg_0 c^\ndg_1} 
+\Gamma_\tau^*\ew{c^\dg_1 c^\ndg_1}\\
\nonumber
&+&i M \ \ew{P^\dg_0 c_1}
-i M \ \ew{c^\dg_0 P^\ndg_1},\\
\partial_t\ew{P^\dg_0 c^\ndg_1}
&=&-\Gamma\ew{P^\dg_0 c^\ndg_1} 
+ i M \ \ew{c^\dg_0 c^\ndg_1}
- i M \ew{P^\dg_0 P^\ndg_1},\\
\partial\ew{c^\dg_0 P^\ndg_1}
&=&-\Gamma\ew{c^\dg_0 P^\ndg_1}
+\Gamma_\tau^*\ew{c_1^\dg P^\ndg_1}\\
\nonumber
&-&i M \ew{c^\dg_0 c^\ndg_1} 
+ i M \ew{P^\dg_0 P^\ndg_1},\\
\partial_t\ew{P^\dg_0 P^\ndg_1}
&=& 
 i M \ \ew{c^\dg_0 P^\ndg_1}
-i M \ \ew{P^\dg_0 c^\ndg_1}
\end{eqnarray}
For this second time interval, there are no additional EOMs required, since
correlations such as $\ew{c^\dg_i c^\ndg_j}$, $\ew{P^\dg_i c^\ndg_j}$ and
$\ew{P^\dg_i P^\ndg_j}$, with $i,j>0$, are already included within 
\eqref{eq:t0_1}-\eqref{eq:t0_3} of the previous time interval. 
With this,
in an arbitrary $\tau$-interval the EOMs thus result in: 
$t_0\in[i\tau,(i+1)\tau)$
\begin{eqnarray}
\nonumber
\partial_t\ew{c^\dg_0 c^\ndg_j}
&=&
-2\Gamma\ew{c^\dg_0 c^\ndg_j} 
+\Gamma_\tau^*\ew{c^\dg_0 c^\ndg_j} 
+ \Gamma_\tau\ew{c^\dg_0 c^\ndg_{j+1}} \Theta_{j+1}\\
\label{eq:general_1}
&+&ig\ew{P^\dg_0 c_j}
-i M \ \ew{c^\dg_0 P^\ndg_j},\\
\partial_t\ew{P^\dg_0 c^\ndg_j}
&=&
-\Gamma\ew{P^\dg_0 c^\ndg_j}
+\Gamma_\tau\ew{P^\dg_0 c^\ndg_{j+1}}\Theta_{j+1}\\
\nonumber
&+& i M \ \ew{c^\dg_0 c^\ndg_j}
-i M \ew{P^\dg_0 P^\ndg_j},\\
\partial\ew{c^\dg_0 P^\ndg_j}
&=&
-\Gamma\ew{c^\dg_0 P^\ndg_j}
+\Gamma_\tau^*\ew{c_1^\dg P^\ndg_j}\\
\nonumber
&-&i M \ \ew{c^\dg_0 c^\ndg_j} 
+ i M \ \ew{P^\dg_0 P^\ndg_j},\\
\partial_t\ew{P^\dg_0 P^\ndg_j}
&=&
i M \ \ew{c^\dg_0 P^\ndg_j}
-i M \ \ew{P^\dg_0 c^\ndg_j}
\label{eq:general_4}
\end{eqnarray}
Here $j$ can be any number from $0$ and $i$. 
At first sight it seems necessary to memorize any possible 
two-time correlation $\ew{A_n B_m}$ in order
to compute this growing set of equations. 
However this is not the case. 
Instead, the number of quantities to memorize grows linearly with the
index $i$ of the time-interval [See Fig. \ref{fig:linear_growth}]. 
Solving the Block of \eqref{eq:general_1} - \eqref{eq:general_4} of the 
$i$-th interval, for any time delay $j\tau\le i\tau$, only two quantities at 
each $j$ from previous times have to be stored:
\begin{eqnarray}
\text{For}\quad j\le i\quad \longrightarrow \quad \{\ew{c^\dagger_1 c^\ndg_j},\
\ew{c^\dagger_1 P^\ndg_j}\}^{\phantom{\dagger}}_j,
\end{eqnarray}
the number of quantities to be stored scales linearly 
with the number of time discretization steps $N_\Delta$ 
as illustrated in Fig. \ref{fig:linear_growth}.
Most of the previous two-time-correlations do not couple into the set of
equations of the $i$th interval, so that the numerical effort as well as the
memory cost is drastically reduced. 
Next to the dynamic memory incorporated within
quantities such as $\ew{c^\dagger_1 c^\ndg_j}$ it is necessary to set initial
conditions at each start of a $\tau$-interval, i.e. at the corners of
the intervals $1\tau$, $2\tau$,...,$i\tau$. 
These initial values, however, are available from the calculations of the 
previous set of equations from the time
interval $(i-1)$. 
These initial values are in particular necessary for feedback
times short compared to the inverse cavity coupling strength $2\pi/g$, i.e. if
there is an overlap between decaying cavity population and fed back photon
population of previous times.

\subsection{Benchmark}
\begin{figure}[t!]
\centering
\includegraphics[width=\figuresize]{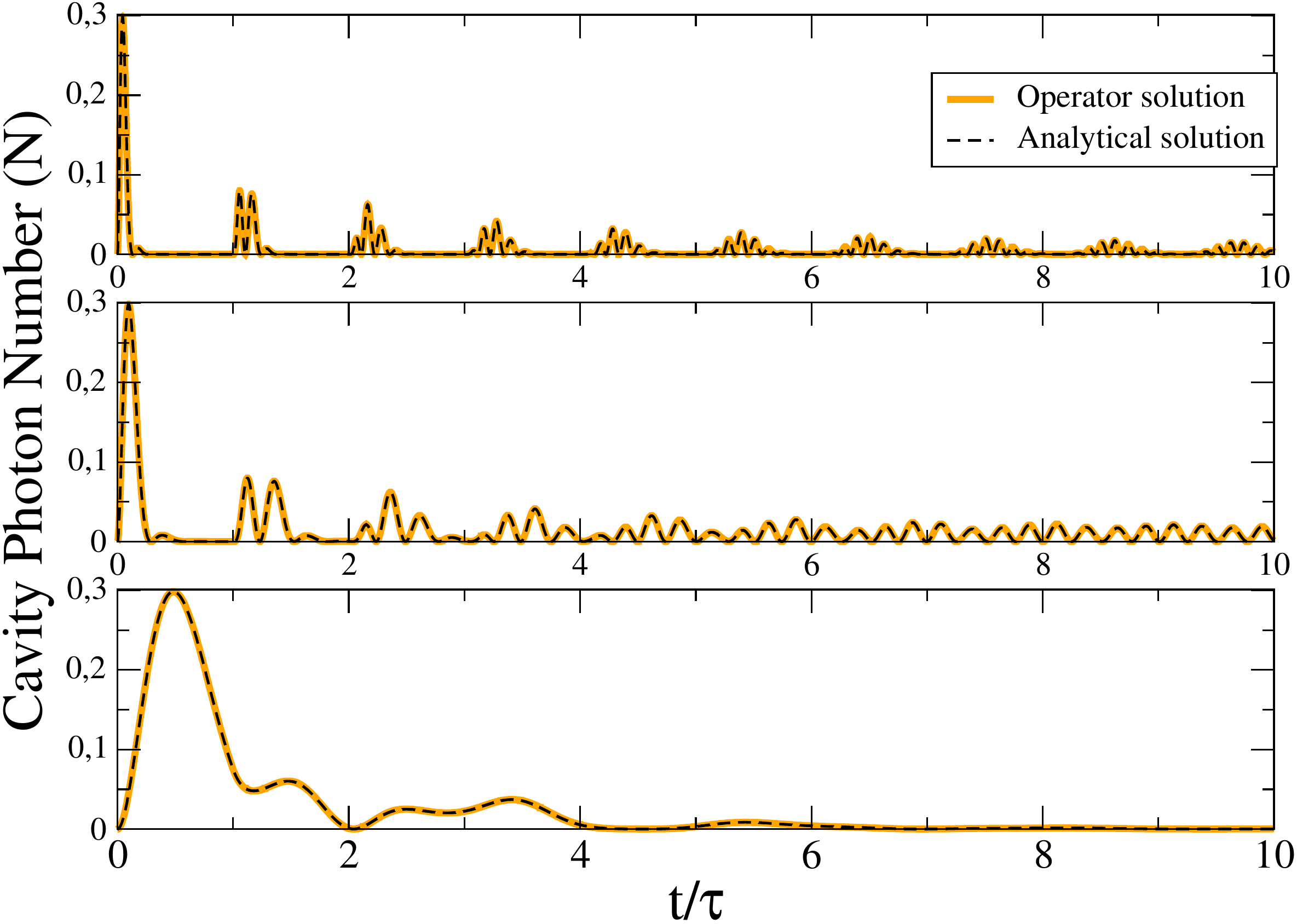}
\caption{Temporal evolution of the cavity photon number: calculated with the
time-correlated operator method (blue curve) and in the Schr\"odinger's picture
(black curve) for three different feedback times $\tau$ and $ M=\Gamma $. (a) 
long feedback
time, (b) feedback time corresponding to a Rabi-oscillation phase and (c) short
feedback time.}
\label{fig:Rabi_comparison}
\end{figure}
In the present section, the temporal evolution of the cavity photon number
[Fig. \ref{fig:Rabi_comparison}] is computed on the basis of 
\eqref{eq:general_1} - \eqref{eq:general_4} (thin blue curves) and as a
benchmark compared with calculations in the Schr\"odingers picture (bold black
curves) \cite{kabuss2015analytical}. 
The evolution is depicted for three different
feedback times $\tau$. 
In Fig. \ref{fig:Rabi_comparison}(a) $\tau$ corresponds
to a rather long feedback time $\tau\gg2\pi/g$. 
Here, the feedback time is so
long, that the entire cavity photon population $\ew{c^\dg_0 c^\ndg_0}$ decays 
into the reservoir, before any population from a previous time is fed back into 
the system. 
In such a case it possible to omit initial values of two-time correlations
$\ew{A(t),B(j\tau)}$ at the corners of the $\tau$-intervals, since these
quantities are zero in that specific case. 
The cavity photon number shows an oscillatory behavior at the time 
scale of the feedback, its envelope
decays completely to zero until the start of the next
$\tau$-interval. 
In the intermediate feedback regime [Fig.
\ref{fig:Rabi_comparison}(b)], with a feedback time corresponding
to $\tau=2\pi/g$ it is possible to stabilize Rabi-oscillations after a series
of round trip times $\tau$ as has been reported in previous works
\cite{carmele2013single,kabuss2015analytical}. 
Here, the feedback time is only visible at the beginning.
After few roundtrips, at about $8\tau$ the cavity photon number shows 
Rabi-oscillations on the time scale of the cavity coupling element, 
mimicking a strong coupling situation at a constant
number of intra cavity excitations. 
However, the strength of the operator
method is best demonstrated in Fig. \ref{fig:Rabi_comparison}(c) at a feedback
time $\tau\ll2\pi/g$, again illustrating a perfect agreement between the two
models. 
In this regime, there is an on-going overlap between in- and outgoing
photon-population. 
Such a situation can only be computed correctly, if the
memory inherited within the time-correlators is regarded adequately.
A great advantage to the Schr\"odinger picture is
naturally provided in the Heisenberg operator
language as now the quantum correlations have become
explicit.
This advantageous scaling property is strongly depended on the
specific system, where a fixed number of excitations is present. 
For driven or pumped systems, the growth of correlation can exceed
easily the linear regime. However, the Heisenberg picture allows 
controlled truncation schemes such as Born factorization, which will 
be discussed in the next section.
%

\subsection{Large Photon Number Limit}
\begin{figure}[t!]
\centering
\includegraphics[width=\figuresize]{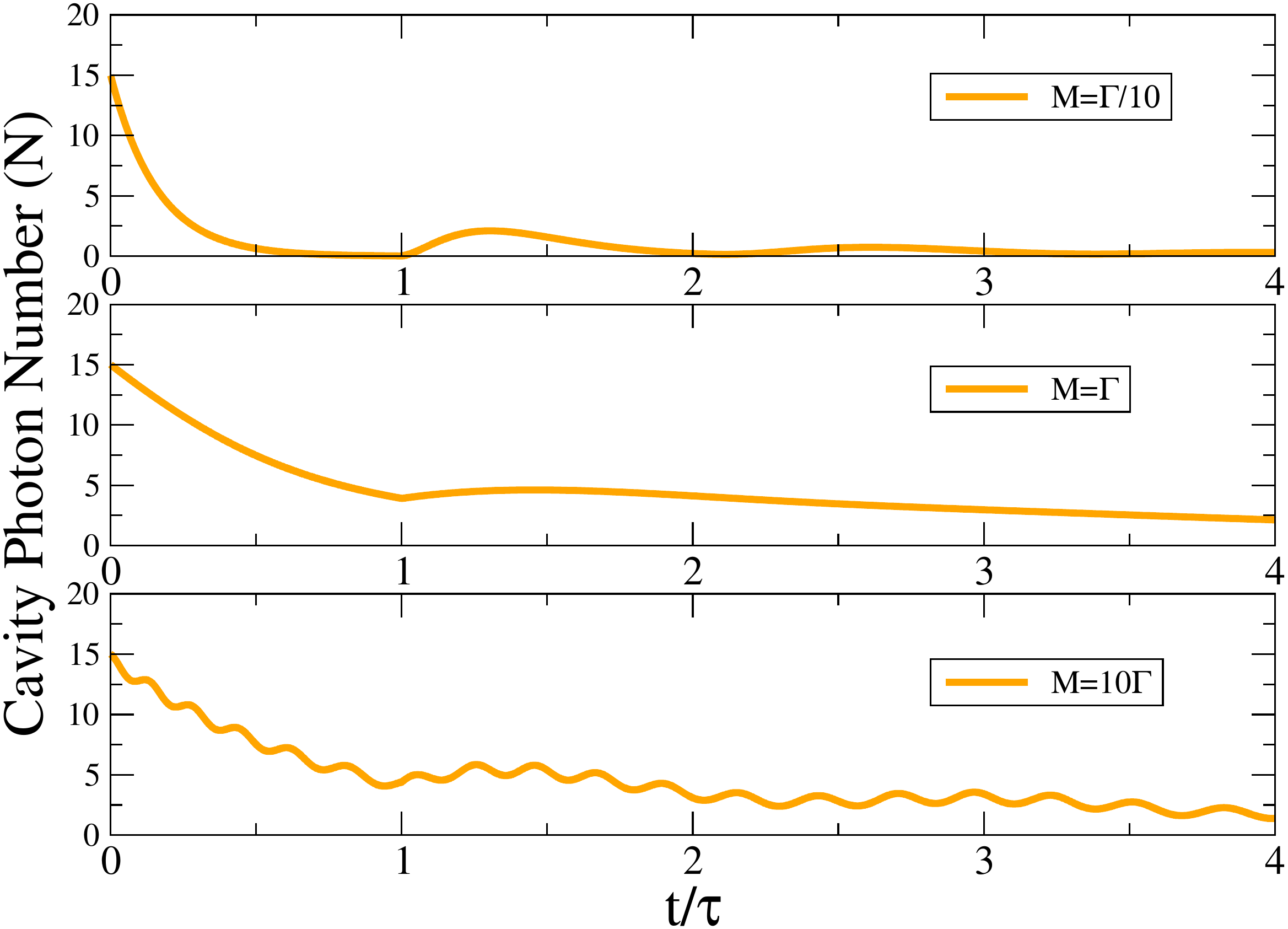}
\caption{Excitation manifold factorization for different ratios of
emitter-cavity strength $ M $ versus the decay and feedback strengh $
\Gamma $. Our factorization approach allows to extract the feedback 
feature in the many-photon regime and gives access to a quantum beating 
corresponding to the coupling strength $ M $. Note, in the upper panel, we 
integrated over a longer absolute time interval to allow for a complete
decay of the cavity photon number.}
\label{fig:factorization}
\end{figure}

The Heisenberg equation of motion approach allows
furthermore to use factorization schemes such as the
Born factorization or cluster expansion techniques
\cite{kira2011semiconductor,schulze2014feedback,kopylov2015dissipative,
lingnau2013feedback}.
In this section, we discuss a factorization 
approach for a scenario where a large number 
of photons is present in the cavity with only
one emitter.
It is known, that a factorization approach is
feasible in this limit \cite{gies2007semiconductor,leymann2014expectation}. 
However, handling delay equations and two-time
correlations, it is still a question how to 
factorize.
Here, we employ a excitation manifold factorization
approach, meaning that we factorize in terms
of an Hilbert space excitation number \cite{richter2009few}.
Investigating the operator dynamics, we pinpoint 
the transition from one excitation manifold
to the next higher in the polarization dynamics:
\begin{eqnarray}
\dot{P}_j&=& 
- i M \ \left(1 -2 P_j^\dagger P_j\right) c_j .
\end{eqnarray}
The polarization exchanges the excitation
within the manifold but couples to the next higher 
manifold via the excited state density.
We factorize therefore between the 
corresponding excitation density and the  
photonic part in the equations of motion, e.g.
\begin{eqnarray}
\ew{\dot{P}_j}
&\approx& 
- i M \ \left(1 -2 \ew{P_j^\dagger P_j}\right) \ew{c_j}.
\end{eqnarray}
This approach is well justified for a large
number of photons and the corresponding set of
equations of motions reads (after factorization):
\begin{eqnarray}
\partial_t\ew{P^\dg_0 c^\ndg_0 } 
&\approx& 
-\Gamma\ew{P^\dg_0 c^\ndg_0 }
+\Gamma_\tau\ew{P^\dg_0 c^\ndg_1}\\
\nonumber
&+& i M \ \ew{c^\dg_0 c^\ndg_0 }
-i M \ \ew{P^\dg_0  P^\ndg_0 } 
-i2M \ \ew{c^\dg_0 c^\ndg_0 } \ew{P^\dg_0  P^\ndg_0},\\
\partial_t\ew{P^\dg_0 c^\ndg_1} \notag
&\approx&
-\Gamma\ew{P^\dg_0 c^\ndg_1} 
-i M \ew{P^\dg_0 P^\ndg_1} \\
&+& i M \ \ew{c^\dg_0 c^\ndg_1}
- i2M \ \ew{c^\dg_0 c^\ndg_1}
\ew{P^\dg_0 P^\ndg_0}, \\
\partial\ew{c^\dg_0 P^\ndg_1}
&\approx&-\Gamma\ew{c^\dg_0 P^\ndg_1}
+\Gamma_\tau^*\ew{c_1^\dg P^\ndg_1}
+ i M \ew{P^\dg_0 P^\ndg_1}\\
\nonumber
&-&i M \ew{c^\dg_0 c^\ndg_1} 
+i2M \ew{c^\dg_0 c^\ndg_1}
\ew{P^\dg_1 P^\ndg_1},\\
\partial_t\ew{P^\dg_0 P^\ndg_1}
&\approx& 
i M \ \ew{c^\dg_0 P^\ndg_1} 
-i2M \ \ew{c^\dg_0 P^\ndg_1} 
\ew{P^\dg_0 P^\ndg_0}\\
&-&i M \ \ew{P^\dg_0 c^\ndg_1}
+i2M \ \ew{P^\dg_0 c^\ndg_1}
\ew{P^\dg_1 P^\ndg_1}
\end{eqnarray}
This set of equation holds only in a regime, where the number
of photons is much larger than the number of emitters.
In Fig.\ref{fig:factorization}, we plotted the dynamics of the 
photon number occupation inside the cavity $ N=\ew{c^\dg c}$ 
with an initial value of $ N=15 $.
In our limit, we clearly see the impact of the feedback.
In the first time interval $[0,\tau]$, the cavity population
decays for all three parameter sets.
Depending on the ratio of $ M/\Gamma $, we see oscillations but
more importantly, we see, that the feedback stops the decay 
and this proportional to the number of excitations.
Note, in Fig.\ref{fig:factorization}(upper panel), we integrated
over a larger time interval to consider a complete decay of the cavity 
population.

\section{Conclusion} \label{sec:conclusion}
We presented a Heisenberg operator method for the description of photon 
feedback in the quantum limit. 
Based on a time-delayed feedback equation for the photon
coherence, that was derived via the elimination of the feedback reservoir,
the dynamics of the system can be calculated by a set of time-correlated
expectations values, and can be computed separately for each time interval.
Here, the memory of the system leading to the photon feedback is calculated on
the fly. 
As the sets of equations corresponding to a certain time interval only
couple to distinctive quantities of the previous and only the previous time
interval, memory carrying time-correlators to be stored just grow for 
the given example linearly with the index of the $\tau$-intervals. 
This results in an extreme reduction
of the numerical effort compared with the incorporation of the reservoir sums,
while still bearing all the memory information. 
The great advantage of the operator method, however, is that it is accessible 
for the description of more complex systems and with multiple excitations and 
different statistical
properties.
As a first outlook, we employed a excitation manifold factorization and showed,
that the Heisenberg operator method paves the way to describe efficient and 
intuitively many-photon quantum feedback.
A next step will include a generalization of the feedback dynamics 
to open, pumped systems.
\section{Acknowledgments}
We would like to thank N. Naumann, M. Kraft and S. Hein for helpful discussions.
We acknowledge support from Deutsche Forschungsgemeinschaft through SFB
910 ``Control of self-organizing nonlinear systems''.


\begin{thebibliography}{10}
\newcommand{\enquote}[1]{``#1''}
\bibitem{zoller-roadmap}
P.~Zoller, T.~Beth, D.~Binosi, R.~Blatt, H.~Briegel \emph{et~al.},
  \enquote{Quantum information processing and communication,} Eur. Phys. J. D
  \textbf{36}, 203--228 (2005).

\bibitem{Wiseman::09}
H.~Wiseman and G.~Milburn, \emph{Quantum Measurement and Control} (Cambridge
  University Press, Oxford, 2006).

\bibitem{brandes2015feedback}
T.~Brandes, \enquote{Feedback between interacting transport channels,} Physical
  Review E \textbf{91}, 052149 (2015).

\bibitem{kiraz2004quantum}
A.~Kiraz, M.~Atat{\"u}re, and A.~Imamo{\u{g}}lu, \enquote{Quantum-dot
  single-photon sources: Prospects for applications in linear optics
  quantum-information processing,} Physical Review A \textbf{69}, 032305
  (2004).

\bibitem{wolters2014deterministic}
J.~Wolters, J.~Kabuss, A.~Knorr, and O.~Benson, \enquote{Deterministic and
  robust entanglement of nitrogen-vacancy centers using low-q photonic-crystal
  cavities,} Physical Review A \textbf{89}, 060303 (2014).

\bibitem{sayrin2011real}
C.~Sayrin, I.~Dotsenko, X.~Zhou, B.~Peaudecerf, T.~Rybarczyk, S.~Gleyzes,
  P.~Rouchon, M.~Mirrahimi, H.~Amini, M.~Brune \emph{et~al.},
  \enquote{Real-time quantum feedback prepares and stabilizes photon number
  states,} Nature \textbf{477}, 73--77 (2011).

\bibitem{dorner-half-cavities}
U.~Dorner and P.~Zoller, \enquote{Laser-driven atoms in half-cavities,} Phys.
  Rev. A \textbf{66}, 023816 (2002).

\bibitem{carmele2013single}
A.~Carmele, J.~Kabuss, F.~Schulze, S.~Reitzenstein, and A.~Knorr,
  \enquote{Single photon delayed feedback: a way to stabilize intrinsic quantum
  cavity electrodynamics,} Physical review letters \textbf{110}, 013601 (2013).

\bibitem{grimsmo2015time}
A.~L. Grimsmo, \enquote{Time-delayed quantum feedback control,} arXiv preprint
  arXiv:1502.06959  (2015).

\bibitem{pichler2015photonic}
H.~Pichler and P.~Zoller, \enquote{Photonic quantum circuits with time delays,}
  arXiv preprint arXiv:1510.04646  (2015).

\bibitem{cook1987quantum}
R.~Cook and P.~Milonni, \enquote{Quantum theory of an atom near partially
  reflecting walls,} Physical Review A \textbf{35}, 5081 (1987).

\bibitem{alber1992photon}
G.~Alber, \enquote{Photon wave packets and spontaneous decay in a cavity,}
  Physical Review A \textbf{46}, R5338 (1992).

\bibitem{tufarelli2014non}
T.~Tufarelli, M.~Kim, and F.~Ciccarello, \enquote{Non-markovianity of a quantum
  emitter in front of a mirror,} Physical Review A \textbf{90}, 012113 (2014).

\bibitem{hein2014optical}
S.~M. Hein, F.~Schulze, A.~Carmele, and A.~Knorr, \enquote{Optical
  feedback-enhanced photon entanglement from a biexciton cascade,} Physical
  review letters \textbf{113}, 027401 (2014).

\bibitem{hughes2007coupled}
S.~Hughes, \enquote{Coupled-cavity qed using planar photonic crystals,}
  Physical review letters \textbf{98}, 083603 (2007).

\bibitem{hetet2011single}
G.~H{\'e}tet, L.~Slodi{\v{c}}ka, M.~Hennrich, and R.~Blatt, \enquote{Single
  atom as a mirror of an optical cavity,} Physical review letters \textbf{107},
  133002 (2011).

\bibitem{albert2011observing}
F.~Albert, C.~Hopfmann, S.~Reitzenstein, C.~Schneider, S.~H{\"o}fling,
  L.~Worschech, M.~Kamp, W.~Kinzel, A.~Forchel, and I.~Kanter,
  \enquote{Observing chaos for quantum-dot microlasers with external feedback,}
  Nature communications \textbf{2}, 366 (2011).

\bibitem{gardiner-book}
C.~Gardiner and P.~Zoller, \emph{Quantum Noise} (Springer, Berlin Heidelberg
  New York, 1991).

\bibitem{kira2011semiconductor}
M.~Kira and S.~W. Koch, \emph{Semiconductor quantum optics} (Cambridge
  University Press, 2011).

\bibitem{schulze2014feedback}
F.~Schulze, B.~Lingnau, S.~M. Hein, A.~Carmele, E.~Sch{\"o}ll, K.~L{\"u}dge,
  and A.~Knorr, \enquote{Feedback-induced steady-state light bunching above the
  lasing threshold,} Physical Review A \textbf{89}, 041801 (2014).

\bibitem{lang1980external}
R.~Lang and K.~Kobayashi, \enquote{External optical feedback effects on
  semiconductor injection laser properties,} Quantum Electronics, IEEE Journal
  of \textbf{16}, 347--355 (1980).

\bibitem{kantner2015delay}
M.~Kantner, E.~Sch{\"o}ll, and S.~Yanchuk, \enquote{Delay-induced patterns in a
  two-dimensional lattice of coupled oscillators,} Scientific reports
  \textbf{5} (2015).

\bibitem{wegert2014integrated}
M.~Wegert, D.~Schwochert, E.~Sch{\"o}ll, and K.~L{\"u}dge, \enquote{Integrated
  quantum-dot laser devices: modulation stability with electro-optic
  modulator,} Optical and Quantum Electronics \textbf{46}, 1337--1344 (2014).

\bibitem{flunkert2013dynamics}
V.~Flunkert, I.~Fischer, and E.~Sch{\"o}ll, \enquote{Dynamics, control and
  information in delay-coupled systems: an overview,} Philosophical
  Transactions of the Royal Society of London A: Mathematical, Physical and
  Engineering Sciences \textbf{371}, 20120465 (2013).

\bibitem{glaetzle2010single}
A.~W. Glaetzle, K.~Hammerer, A.~Daley, R.~Blatt, and P.~Zoller, \enquote{A
  single trapped atom in front of an oscillating mirror,} Optics Communications
  \textbf{283}, 758--765 (2010).

\bibitem{lei2012quantum}
C.~U. Lei and W.-M. Zhang, \enquote{A quantum photonic dissipative transport
  theory,} Annals of Physics \textbf{327}, 1408--1433 (2012).

\bibitem{kabuss2015analytical}
J.~Kabuss, D.~O. Krimer, S.~Rotter, K.~Stannigel, A.~Knorr, and A.~Carmele,
  \enquote{Analytical study of quantum feedback enhanced rabi oscillations,}
  arXiv preprint arXiv:1503.05722  (2015).

\bibitem{flunkert2011time}
V.~Flunkert, \enquote{Time-delayed feedback control,} in \enquote{Delay-Coupled
  Complex Systems,}  (Springer, 2011), pp. 7--10.

\bibitem{eschner2001light}
J.~Eschner, C.~Raab, F.~Schmidt-Kaler, and R.~Blatt, \enquote{Light
  interference from single atoms and their mirror images,} Nature \textbf{413},
  495--498 (2001).

\bibitem{hein2015entanglement}
S.~M. Hein, F.~Schulze, A.~Carmele, and A.~Knorr, \enquote{Entanglement control
  in quantum networks by quantum-coherent time-delayed feedback,} Physical
  Review A \textbf{91}, 052321 (2015).

\bibitem{carmele2014opto}
A.~Carmele, B.~Vogell, K.~Stannigel, and P.~Zoller, \enquote{Opto-nanomechanics
  strongly coupled to a rydberg superatom: coherent versus incoherent
  dynamics,} New Journal of Physics \textbf{16}, 063042 (2014).

\bibitem{kopylov2015dissipative}
W.~Kopylov, M.~Radonji{\'c}, T.~Brandes, A.~Bala{\v{z}}, and A.~Pelster,
  \enquote{Dissipative two-mode tavis-cummings model with time-delayed feedback
  control,} arXiv preprint arXiv:1507.01811  (2015).

\bibitem{lingnau2013feedback}
B.~Lingnau, W.~W. Chow, E.~Sch{\"o}ll, and K.~L{\"u}dge, \enquote{Feedback and
  injection locking instabilities in quantum-dot lasers: a microscopically
  based bifurcation analysis,} New Journal of Physics \textbf{15}, 093031
  (2013).

\bibitem{gies2007semiconductor}
C.~Gies, J.~Wiersig, M.~Lorke, and F.~Jahnke, \enquote{Semiconductor model for
  quantum-dot-based microcavity lasers,} Physical Review A \textbf{75}, 013803
  (2007).

\bibitem{leymann2014expectation}
H.~Leymann, A.~Foerster, and J.~Wiersig, \enquote{Expectation value based
  equation-of-motion approach for open quantum systems: A general formalism,}
  Physical Review B \textbf{89}, 085308 (2014).

\bibitem{richter2009few}
M.~Richter, A.~Carmele, A.~Sitek, and A.~Knorr, \enquote{Few-photon model of
  the optical emission of semiconductor quantum dots,} Physical review letters
  \textbf{103}, 087407 (2009).

\end{thebibliography}
\end{document}